\documentclass[12pt]{iopart}
\usepackage{iopams}
\usepackage{graphicx}
\usepackage{color}
\usepackage{amssymb}

\begin{document}

\title[Quench dynamics of the Tomonaga-Luttinger model]
{Quench dynamics of the Tomonaga-Luttinger model with momentum dependent 
interaction}

\author{J Rentrop, D Schuricht and V Meden}

\address{Institut f\"ur Theorie der Statistischen Physik, RWTH Aachen University 
and JARA---Fundamentals of Future Information Technology, 52056 Aachen, Germany}

\ead{meden@physik.rwth-aachen.de}

\begin{abstract}
We study the relaxation dynamics of the one-dimensional Tomonaga-Luttinger model 
after an interaction quench paying particular attention to the momentum 
dependence of the two-particle interaction. Several potentials of different 
analytical form are investigated all leading to universal Luttinger liquid 
physics in equilibrium. The steady-state fermionic momentum 
distribution shows universal behavior in the sense of the Luttinger liquid 
phenomenology. For generic regular potentials the large time decay of 
the momentum distribution function towards the steady-state value 
is characterized by a power law with a universal 
exponent which only depends  on the potential 
at zero momentum transfer. A commonly employed ad hoc procedure 
fails to give this exponent. Besides quenches from zero 
to positive interactions we also consider 
abrupt changes of the interaction between two arbitrary values. 
Additionally, we discuss the appearance of a factor of two 
between the steady-state momentum distribution function and the one 
obtained in equilibrium at equal two-particle interaction.    
\end{abstract}

\maketitle

\section{Introduction}
\label{sec:introduction}

The experimental progress in controlling and manipulating cold atomic 
gases \cite{Bloch08}, which in certain parameter regimes form strongly 
correlated quantum many-body systems, led to tremendous activities 
with the final goal of theoretically understanding the quench dynamics 
in such systems (for a recent review, see Ref.~\cite{Polkovnikov11}). In a 
sudden quench at least one parameter of a given Hamiltonian is switched 
abruptly at time $t=0$---the 
Hamiltonian before the quench is $H_{\rm i}$, the one afterwards $H_{\rm f}$. At $t=0$ 
the system is assumed to be in the ground state (we consider temperature $T=0$)
of $H_{\rm i}$ and the time evolution for $t>0$ is performed with
respect to $H_{\rm f}$. We here study 'global' quenches of the
two-particle interaction.     

Describing the time evolution of a correlated quantum system out of a nonequilibrium state poses 
a formidable challenge. It is reasonable to consider the case of one-dimensional (1d) 
systems first, as in 1d a variety of analytical as well as numerical methods exist which allow for 
controlled access to equilibrium correlation effects in specific models 
\cite{Voit95,Giamarchi03,Schoenhammer05,Essler05}. Several of those techniques were recently 
extended to study the nonequilibrium problem at hand 
\cite{Daley04,Manmana07,Karrasch11,Calabrese07,Kollar08,Schiro10,Mossel10,Gritsev10,Calabrese11,Goth11}. 
Furthermore, in 1d chains virtually all 
many-body states of matter of current interest, such as Mott insulators, superfluids, 
superconductors and charge- as well as spin-density-wave states can be
realized. 
Quenching between these states is expected to be particularly 
interesting. 

We here focus on fermionic systems. Even if the system stays metallic in the presence 
of two-particle interactions the Fermi liquid concept breaks down. Instead a wide class of 
1d models shows Luttinger liquid (LL) behavior on low-energy scales 
\cite{Voit95,Giamarchi03,Schoenhammer05}. Equilibrium LL physics is characterized by 
the 'universal' power-law decay of certain correlation functions in space-time with 
exponents which for spinless and spin-rotational invariant models 
can all be expressed in terms of a single number $K$. This LL parameter in turn is a function
of the microscopic details of the model considered, in particular the 
strength of the two-particle interaction. For noninteracting fermions $K=1$ and
$0< K <1$ for repulsive ones; the case we focus on. 
    
The Tomonaga-Luttinger (TL) model is the exactly solvable effective low-energy 
fixed point model of the LL 
universality class \cite{Haldane81,Solyom79,Voit95,Giamarchi03,Schoenhammer05}. 
It thus plays a similar role as the free Fermi gas in Fermi liquid theory. The model
has two strictly linear branches of right- and left-moving fermions and two-particle
scattering is restricted to processes with small momentum transfer $|q| \ll k_{\rm F}$, with
the Fermi momentum $k_{\rm F}$. These processes as well as the kinetic energy can be written 
as quadratic forms of the densities of right- and left-moving fermions which obey bosonic 
commutation relations. In most calculations the momentum dependence 
of the low-momentum scattering processes $g_2$ and $g_4$ (in the so-called 
g-ology classification \cite{Solyom79}; see below) are (partially) neglected and momentum 
integrals are regularized in the ultraviolet introducing a convenient 
cutoff 'by hand'.

In the bosonization procedure the interacting Fermi system is mapped onto a model
of free bosons. The time evolution of the latter is trivial and the
dynamics of correlation functions of the fermionic densities 
(at small momenta) after an interaction quench can be
accessed directly. 
In addition, the fermionic field operator can be written as a 
(highly nonlinear) function of the bosonic eigenmodes $\alpha_n^{(\dag)}$
\cite{Haldane81,Voit95,Giamarchi03,Schoenhammer05} such that the time evolution of fermionic 
correlation functions can be computed exactly as well. The TL model thus constitutes 
an ideal play ground for studying the dynamics  resulting from an 
interaction quench. 

The time evolution of the fermionic single-particle Green function $
G_t(x) = \left< \psi^\dag(x) \psi(0) \right>_{\rho(t)}$ as well as the 
density correlation function (at small momenta) after suddenly turning on 
the interaction in the spinless TL model 
was first studied in Ref.~\cite{Cazalilla06}. Here $\psi^{(\dag)}(x)$ denotes the 
field operator and  
$ \left< \ldots \right>_{\rho(t)}$ the expectation value with respect to the time-dependent 
density matrix $\rho(t)$. In  Ref.~\cite{Cazalilla06} it was found that at large $t$ 
the Green function approaches a time-independent stationary limit.  At $T=0$ 
the stationary Green function shows power-law behavior as a function of the position 
$x$ with a $K$ dependent exponent. Power-law decay of 
$\lim_{t \to \infty} G_t(x) = G^{\rm st}(x)$ at large $|x|$ translates into a typical LL 
power-law behavior of the stationary fermionic momentum distribution function 
close to $k_{\rm F}$, 
$|n^{\rm st}(k) -1/2| \sim |k-k_{\rm F}|^{\gamma_{\rm st}}$, with $\gamma_{\rm st}$ 
being a function of $K$ and thus depending on the strength of the interaction.
Interestingly, the steady-state exponent $\gamma_{\rm st}$ differs from the 
one of the ground-state momentum distribution function $\gamma_{\rm gs}$ 
at the same interaction strength. For finite times $n(k,t)$ has a Fermi 
liquid-like jump at $k_{\rm F}$, with a $Z$-factor which vanishes as a power law 
in $t$: $Z \sim t^{-\gamma_{\rm st}}$. Further aspects of the quench dynamics of 
the TL model or closely related ones were discussed in 
Refs.~\cite{Perfetto06,Kennes10,Dora11,Mitra11,Iucci09}.

In Ref.~\cite{Cazalilla06} an ad hoc ultraviolet regularization was used. 
It is widely believed that (partially) neglecting the momentum dependence
of the interaction and regularizing momentum integrals as convenient has 
no effect on the low-energy {\em equilibrium} physics of the TL model. This is indeed 
correct if all energy scales are sent to zero \cite{Meden99}. The such obtained 
results for the dependence of power-law exponents of 
specific correlation functions on the LL parameter $K$ become universal 
and are valid for all models falling into the LL universality class. 
It is however questionable if the same reasoning holds when considering quenches. 
High energy processes and thus the full momentum dependence of the 
two-particle interaction might matter in this nonequilibrium situation. 
To investigate this issue we keep the momentum dependence of the 
$g_2$ and $g_4$ processes---rendering any 
ad hoc ultraviolet regularization superfluous 
\cite{Meden99}---and consider interaction potentials of different analytical 
form. We exactly compute $G_t(x)$ and its Fourier transform, the fermionic momentum 
distribution function $n(k,t)$, of the spinless TL model after a quench out of the 
noninteracting ground state. We first show that independent 
of the details of the momentum dependence of the interaction the {\em steady-state} 
momentum distribution as a function of $k$ is characterized by a 
power-law nonanalyticity at $k_{\rm F}$ with the exponent $\gamma_{\rm st}(K)$.
The LL parameter $K$ is a function of the potential at zero  momentum transfer 
only. Therefore $n^{\rm st}(k)$ of the TL model is {\it universal} in the LL 
sense. Similarly, the $Z$-factor shows a universal power-law decay in 
time. We then proceed and show that 
for generic regular potentials also the asymptotic {\em time dependence} of $n(k,t)$
has {\em universal} aspects. For fixed $k \neq k_{\rm F}$ we find a power-law
decay towards the steady-state value as a function of $t$ with the $k$-independent 
exponent $1+ \gamma_{\rm st}$. The ad hoc procedure instead gives the 
exponent $1+3\gamma_{\rm st}/2$ and is thus insufficient for studies of the 
generic dynamics. The power-law decay is overlaid by an oscillation with a frequency 
which depends on the momentum $k$ considered as well as the momentum dependence of 
the potential. In addition, we investigate a box shaped (in momentum space) 
potential and show that the asymptotic dynamics is dominated by 
its discontinuity leading to a 
power-law decay with exponent $1$ (independent of 
the interaction strenght).  
We raise the question if the universality in the steady state as well as the large time 
dynamics of the TL model after an interaction quench extend to other models of the 
(equilibrium) LL universality class. Besides quenches from zero to positive interactions 
we briefly consider quenches between two interactions of arbitrary strength. 

Our paper is organized as follows. In Sec.~\ref{modbos} we present the model and give 
details on how to compute the single-particle Green function after 
a quench. Particular emphasis is put on the ultraviolet 
regularization by momentum dependent two-particle potentials. Different model 
potentials are introduced. In Sec.~\ref{ssee} we compute and compare the 
steady-state and equilibrium Green and momentum distribution functions. 
For small interactions they differ by a factor of two which was earlier discussed 
in the context of pre-thermalized states \cite{Moeckel08,Moeckel09,Kollar11}.
In Sec.~\ref{timeevolution} we present our results for $n(k,t)$ and show 
that they partly depend on the form of the potential considered. We also 
discuss results for the standard ad hoc regularization.  Finally, our 
findings are summarized in Sec.~\ref{summary}.  

\section{Model and methods} 
\label{modbos}

\subsection{The Tomonaga-Luttinger model and bosonization}

The interaction energy of a translational invariant system (periodic boundary conditions) of 1d 
spinless fermions interacting via a potential which only depends on the distance between the two 
scattering particles can be written as 
\begin{eqnarray}  
H_{\rm pot} = \frac{1}{L} \, \sum_{n>0} v(q_n) \rho_n \rho_{-n} + h_1(N) \; , \;\;\;  q_n = \frac{2 \pi}{L} \, n  \; , \;\;\;  
n \in {\mathbb Z} \; ,  
\label{Hpotdef}
\end{eqnarray}
with the density operator 
\begin{eqnarray} 
\rho_n = \sum_{m} c_m^\dag c_{m+n}^{}  
\label{densitydef}
\end{eqnarray} 
and the Fourier transform $v(k)$ of the two-particle potential $V(x)$. Here the $c_n^{(\dag)}$  denote 
fermionic momentum space annihilation (creation) operators, $h_1(N)$ contains terms which depend on the 
particle number operator $N$, and $L$ is the length of the system. After linearization of the 
single-particle dispersion around the two Fermi points at $\pm k_{\rm F}$ the kinetic energy reads
\begin{eqnarray}
\mbox{} \hspace{-1cm} H_{\rm kin} = \sum_{n>0} v_{\rm F} (k_n-k_{\rm F}) c_{n,+}^\dag c_{n,+}^{} + 
 \sum_{n<0} (-v_{\rm F}) (k_n+k_{\rm F}) c_{n,-}^\dag c_{n,-}^{} + h_2(N) \; ,
\end{eqnarray}  
where we already introduced independent right- ($c_{n>0,+}$) and left-moving ($c_{n<0,-}$) 
fermions \cite{Schoenhammer05}. The Fermi velocity is denoted by $v_{\rm F}$. 
In the next step one supplements the Hilbert space of the right movers by states with negative 
momenta and the one of left movers by states with positive momenta. The linearization and 
addition of states does not change the 
equilibrium low-energy physics. 
From now on we drop  terms containing $N$ as they are irrelevant for our considerations. 
The kinetic energy $H_{\rm kin}$ 
can then be written as a quadratic form in the densities  $\rho_{n,\pm}$ of the right and left movers 
defined in analogy  to Eq.~(\ref{densitydef}) \cite{Schoenhammer05}. With
\begin{eqnarray}
b_n = \frac{1}{\sqrt{|n|}} \left\{ \begin{array}{cc} 
\rho_{n,+} & \mbox{for} \; n >0 \\
\rho_{n,-} & \mbox{for} \; n < 0 
\end{array} \right. \; ,
\label{bdef}
\end{eqnarray} 
one obtains
\begin{eqnarray}
H_{\rm kin} = \sum_{n \neq 0} v_{\rm F} |k_n| b_n^\dag b_n^{} \; .
\label{Hkindef}
\end{eqnarray} 
The $b_n^{(\dag)}$ obey the standard bosonic commutation relations. Replacing 
$\rho_n \to \rho_{n,+} + \rho_{n,-}$ in Eq.~(\ref{Hpotdef}), using Eq.~(\ref{bdef})  
we obtain the Hamiltonian of the TL model
\begin{eqnarray} 
\mbox{} \hspace{-2cm} H_{\rm TL} = \sum_{n >0 } \left[ k_n \left( v_{\rm F} + \frac{v(k_n)}{2 \pi} \right) 
\left( b_n^\dag b_n^{} + b_{-n}^\dag b_{-n}^{} \right) +   k_n   \frac{v(k_n)}{2 \pi} 
\left( b_n^\dag b_{-n}^\dag + b_{-n}^{} b_{n}^{} \right)  \right] \; .
\end{eqnarray}
Distinguishing between intra- and inter-branch scattering processes one often replaces the potential 
$v(k)$ in the first term by a function $g_4(k)$ and the one in the second by an independent function
$g_2(k)$ \cite{Solyom79}. For simplicity we here refrain from doing so and
assume $g_2(k) = g_4(k) = v(k)$. This has no effect on our main results.      

\subsection{Eigenmodes and dynamics}

By a Bogoliubov transformation the 'bosonized' Hamiltonian $H_{\rm TL}$ can straightforwardly 
be diagonalized 
\begin{eqnarray}
H_{\rm TL} = \sum_{n \neq 0} \omega(k_n) \, \alpha_n^\dag \alpha_n^{}  +
E_{\rm gs}
\label{HTLdef}
\end{eqnarray} 
by introducing the eigenmodes
\begin{eqnarray} 
\alpha_n^{} =  c(k_n) b_n^{} - s(k_n) b_{-n}^\dag  \;\; \Leftrightarrow \;\; b_n^{} = c(k_n) \alpha_n^{} 
+ s(k_n) \alpha_{-n}^\dag 
\label{bogo}
\end{eqnarray} 
with
\begin{eqnarray} 
&& s^2(k_n) = \frac{1}{2} \left[ \frac{1+ \hat v(k_n) /2 }{\sqrt{1+ \hat v(k_n)}} -1  \right]  \; , \;\; 
c^2(k_n) = 1+ s^2(k_n)  \; , \nonumber \\  && 
\omega(k_n) = v_{\rm F} \,  |k_n| \, \sqrt{1+\hat v(k_n)} \; , \;\; \hat v(k_n) = \frac{v(k_n)}{\pi v_{\rm F}} \; . 
\label{manydefs}
\end{eqnarray} 
In the noninteracting limit $s^2(k_n) \to 0$ and $\omega(k_n) \to v_{\rm F} |k_n|$. 
For physical reasons the Fourier transform $v(q)$ of the two-particle potential
must vanish on a characteristic scale denoted by $q_{\rm c}$. This implies  
\begin{eqnarray}
\lim _{k \to \infty} s^2(k) = 0 \; , \;\;\;\;\; \lim _{k \to \infty} c^2(k) = 1  
\; , \;\;\;\;\; \lim _{k \to \infty} \frac{\omega(k)}{|k|} =  v_{\rm F}  \; . 
\label{largemomenta}
\end{eqnarray}
The LL parameter $K$ of the TL model is
\begin{eqnarray}
K = \left[1+ \hat v(0) \right]^{-1/2} \; .
\label{LLpardef}
\end{eqnarray}  
It thus only depends on the (dimensionless; see Eq.~(\ref{manydefs})) potential at momentum $q=0$. 
This shows that the TL model is a LL with $0<K<1$ only if $\hat v(0) > 0$; we here focus
on two-particle potentials with this property.  
Using Eq.~(\ref{manydefs}) one finds 
\begin{eqnarray}
s^2(0) = \frac{1}{4} \, \left( K+K^{-1} -2 \right) \; , \; \;\; 
c^2(0) = \frac{1}{4} \, \left( K+K^{-1} +2 \right) \; .
\label{scK}
\end{eqnarray}   
It turns out to be useful to introduce a renormalized velocity $\tilde v_{\rm F}$ and its 
dimensionless analog $\hat v_{\rm F}$ 
\begin{eqnarray}
\tilde v_{\rm F} = \left. \frac{d \omega(k)}{d k} \right|_{k=0} \; , \;\;\; 
\hat v_{\rm F} = \frac{\tilde v_{\rm F}}{v_{\rm F}} \; .
\label{hatvF}
\end{eqnarray}
Using Eq.~(\ref{manydefs}) we find 
\begin{eqnarray}
\hat v_{\rm F} = \sqrt{1+ \hat v(0)} \; .
\label{renvfdef}
\end{eqnarray}
 
The ground state of $H_{\rm TL}$ is given by the vacuum with respect to the eigenmodes 
$\left| \mbox{vac}(\alpha) \right>$ and 
\begin{eqnarray} 
E_{\rm gs} = - 2 v_{\rm F} \sum_{n>0}  k_n \, s^2(k_n) \, \sqrt{1+ \hat v(k_n)}
\label{eee}
\end{eqnarray}  
is the ground-state energy. For vanishing two-particle interaction $E_{\rm gs}^0 = 0$. 

The time evolution with respect to $H_{\rm f} = H_{\rm TL} $ of the eigenmode 
annihilation and creation operators in the Heisenberg 
picture is now trivially given by 
\begin{eqnarray} 
\alpha_n^{} (t) = e^{-i \omega(k_n) t} \, \alpha_n \; , \;\;\; \alpha_n^\dag
(t) = e^{i \omega(k_n) t} \alpha_n^\dag
\label{modetime}
\end{eqnarray}
and the one of the $b_n^{(\dag)}$ can directly be obtained using this and Eq.~(\ref{bogo}).   

\subsection{Bosonization of the field operator and time evolution of the Green function}

To obtain expectation values of fermionic operators, such as the momentum distribution 
function of the right movers (the left movers can be treated similarly) 
\begin{eqnarray} 
n(k_n,t)= \int_{-L/2}^{L/2} dx \, e^{i k_n x} \, G_t(x) \; , \;\;\; G_t(x) =  \left< \psi_+^\dag(x) \psi_+(0)^{} \right>_{\rho(t)} \; ,
\label{nofk}
\end{eqnarray}  
one has to 'bosonize' the fermionic field operator of the right-moving particles     
\begin{equation}
\label{field}
\psi_+^{\dagger}(x)= \frac{1}{\sqrt{L}} \sum_{n} e^{-ik_nx}  c_{n,+}^{\dagger} \;\; .
\end{equation} 
One can prove the operator identity \cite{Haldane81,Voit95,Giamarchi03,Schoenhammer05}
\begin{equation}
\label{bosonization1}
\psi^{\dagger}_+(x)= \frac{e^{- i x\pi/L}}{\sqrt{L}} 
e^{-i \Phi^{\dagger}(x)} U^{\dagger} e^{-i \Phi(x)} \;\; ,
\end{equation}
with
\begin{equation}
\label{bosonization2}
\Phi(x)=\frac{\pi}{L} N  x -i \sum_{n>0} e^{iq_nx} \left(
\frac{2\pi}{L q_n} \right)^{1/2}  b_n \;\; ,
\end{equation}
where $U^{\dagger}$ denotes a unitary fermionic raising operator 
which commutes with the $ b_n^{(\dag)}$ and maps
the $N$-electron ground state to the $(N+1)$-electron one. 
In the computation of the Green function the fermionic operators 
lead to the phase factor $\exp{(-i k_{\rm F} x)}$ appearing below (see 
Refs.~\cite{Schoenhammer05} and \cite{Meden99} for details). 

With the initial density matrix $\rho_{\rm i} =\left| \mbox{vac}(b) \right>
\left< \mbox{vac}(b) \right|$ corresponding to  the noninteracting ground state 
and the time evolution given by the interacting
Hamiltonian Eq.~(\ref{HTLdef}) we obtain
\begin{eqnarray}
&& \mbox{} \hspace{-2.5cm} G_t(x) 
=   \left<  \psi_+^\dag(x) \psi_+(0)^{}  \right>_{\rho(t)} =
\left< \mbox{vac}(b) \right|  
\psi_+^\dag(x,t) \psi_+(0,t)^{} \left| \mbox{vac}(b) \right>
\nonumber \\
&& \mbox{} \hspace{-2.cm}   =  \frac{1}{L} \, \frac{e^{-i k_{\rm F} x}}{1-e^{i\left( \frac{ 2 \pi x +i0}{L}\right)}} 
\exp{ \left\{ \sum_{n>0} \frac{4 s^2(k_n) c^2(k_n)}{n} \left( \cos{[k_n x]} -1
    \right) \left( 1- \cos{\left[2 \omega(k_n) t\right]} \right)  \right\} }  \; ,
\end{eqnarray}
where we used Eqs.~(\ref{bogo}) and (\ref{modetime}) as well as the
Baker-Hausdorff relation. To prevent recurrence 
effects at large times we take the thermodynamic limit 
\begin{eqnarray}
\mbox{} \hspace{-2cm} G_t(x) = \frac{i}{2 \pi} \, 
\frac{e^{-i k_{\rm F} x}}{x +i0} 
\exp{ \left\{ \int_0^\infty  \!\!\! dk \, \frac{4 s^2(k) c^2(k)}{k} \left( \cos{[k x]} -1
    \right) \left(  1- \cos{\left[2 \omega(k) t\right]} \right)  \right\} } \; .
\label{GTL}
\end{eqnarray}
We stress that because of Eq.~(\ref{largemomenta}) the momentum integral is convergent at large 
$k$ and does not require any regularization. 

A more general situation arises if one starts at $t=0$ in the ground state 
with two-particle potential $v_{\rm i}(q)$ and performs the time evolution 
in the presence of the potential $v_{\rm f}(q)$ (an analogous situation for
bosonic LLs is discussed in Ref.~\cite{Mitra12}). Applying the two Bogoliubov
transformations from the noninteracting eigenmodes to the ones with 
$v_{\rm i}(q)$ and $v_{\rm f}(q)$ given by relations analogous to Eqs.~(\ref{bogo}) 
and (\ref{manydefs}) one can generalize Eq.~(\ref{GTL}) to (in self-explaining 
notation)
\begin{eqnarray}
\mbox{} \hspace{-2cm} && G_t(x) =   \frac{i}{2 \pi} \, 
\frac{e^{-i k_{\rm F} x}}{x +i0} \exp  \left\{ 2 \int_0^\infty  \!\!\! dk \, \frac{\cos{[k x]} -1 }{k} 
\right. \nonumber \\* 
\mbox{} \hspace{-2cm} && 
\times \left( c_{\rm f}^2(k) \left[ c_{\rm f}(k) s_{\rm i}(k) - s_{\rm f}(k) c_{\rm i}(k)
\right]^2 + s_{\rm f}^2(k) \left[ c_{\rm f}(k) c_{\rm i}(k) - s_{\rm f}(k) s_{\rm i}(k)
\right]^2 \phantom{\frac{1}{2}}\right.  \nonumber \\*
\mbox{} \hspace{-2.cm} && \hspace{.3cm} \left. \left. +  2 s_{\rm f}(k) c_{\rm f}(k) \left[ c_{\rm f}(k) s_{\rm i}(k) - s_{\rm f}(k) c_{\rm i}(k)
\right] \left[ c_{\rm f}(k) c_{\rm i}(k) - s_{\rm f}(k) s_{\rm i}(k)
\right] \cos{ \left[ 2 \omega_{\rm f}(k) t  \right]} 
\phantom{\int \frac{1}{2}} \hspace{-.8cm} \right) \right\}  \, .
\label{GTLWW1WW2}
\end{eqnarray}
For $v_{\rm i}(q)=0$, implying 
$c_{\rm i}(k)=1$ and $s_{\rm i}(k)=0$, this expression reduces to Eq.~(\ref{GTL}). 

Equation (\ref{GTLWW1WW2}) also covers the interesting situation in which one 
considers the ground state with nonvanishing $v_{\rm i}(q)$ as the initial state 
and performs the time evolution with the {\em noninteracting} Hamiltonian, that is 
for $v_{\rm f}(q)=0$ and thus $c_{\rm f}(k)=1$ and $s_{\rm f}(k)=0$ \cite{Goth11}.
In this case the {\em fermionic} momentum occupancy is conserved. Accordingly 
the time dependence in Eq.~(\ref{GTLWW1WW2}) drops out and the Green function reads
\begin{eqnarray}
G_t(x) = \frac{i}{2 \pi} \, 
\frac{e^{-i k_{\rm F} x}}{x +i0} 
\exp{ \left\{ \int_0^\infty  \!\!\! dk \, 2 s_{\rm i}^2(k) \, \frac{ \cos{[k x]} -1 }{k} 
\right\} } = G^{\rm gs}(x)\; .
\label{GTLequ}
\end{eqnarray} 
It corresponds to the {\em equilibrium} ground-state Green function $G^{\rm gs}(x)$ 
of the TL model at interaction $v_{\rm i}(q)$. Although the time evolution is 
with a noninteracting Hamiltonian 
the system is characterized by the LL Green (and momentum distribution) function of the
initial state for all times. 

\subsection{Momentum integrals and potentials}

Computing expectation values in the TL model in and out of equilibrium one regularly 
faces momentum integrals of
the type appearing in Eqs.~(\ref{GTL}), (\ref{GTLWW1WW2}), and (\ref{GTLequ})
\cite{Voit95,Giamarchi03,Schoenhammer05,Meden99,Cazalilla06,Kennes10}. To evaluate those
one often {\em assumes} that the prefactor of the trigonometric functions has a
convenient exponential form. In the case of the time evolution (under the
interacting Hamiltonian) out of the noninteracting ground state one sets
(see Eq.~(\ref{GTL}))\footnote{In ground-state calculations one instead sets 
$ 2 s_{\rm i}^2(q) =  g^2 \exp{(-|q|/q_{\rm c})}$ \cite{Luther74}, see Eq.~(\ref{GTLequ}).}
\begin{eqnarray} 
4 s^2(q) c^2(q) = g^2 e^{-|q|/q_{\rm c}} \; .
\end{eqnarray} 
Using Eq.~(\ref{manydefs}) one can solve for the momentum dependence of the
underlying two-particle potential and obtains 
\begin{eqnarray} 
\hat v(q) = 2 g^2  e^{-|q|/q_{\rm c}} + 2 \sqrt{ g^2  e^{-|q|/q_{\rm c}} } \, \sqrt{ 1+ g^2
  e^{-|q|/q_{\rm c }}} \; .
\label{pfusch}
\end{eqnarray}  
The momentum-space potential decays exponentially for $|q|/q_{\rm c} \gg 1$. Even with this special choice
of the momentum dependence of the two-particle potential the integral in  
Eq.~(\ref{GTL}) cannot be solved analytically as $\omega(k)$ appears in the 
argument of the cosine. One therefore {\em linearizes} the dispersion and makes the replacement 
\begin{eqnarray} 
\omega(k) \rightarrow \tilde v_{\rm F} |k| \;\; \forall k \; ,
\label{pfusch_2}
\end{eqnarray}  
with $ \tilde v_{\rm F}$ given in Eq.~(\ref{hatvF}).
Equations (\ref{pfusch}) and (\ref{pfusch_2}) form one of the possible {\em ad hoc procedures} 
used in the literature \cite{Voit95,Giamarchi03,Schoenhammer05,Meden99,Cazalilla06} to 
make analytical progress; others are applied as well. With these a closed analytical 
expression for Eq.~(\ref{GTL}) can be given (see Sec.~\ref{timeevolution}).
It is usually believed that this replacement (and similar ones) does not change the 
equilibrium low-energy 
physics, which is indeed correct if all energy scales are sent to zero
\cite{Meden99}.
However, in nonequilibrium the dynamics is affected by the high-energy modes 
and thus the full momentum dependence of the potential becomes important.
The replacement was still used  in Refs.~\cite{Cazalilla06,Iucci09} based 
on the expectation that at least in the weak coupling limit it will
not significantly affect the time dependence.
Here we do not rely on this approximation and exactly evaluate the  
integral in  Eq.~(\ref{GTL}) for the potentials 
\begin{eqnarray} 
v_{\rm box}(q) & = & \left\{ \begin{array}{cc} v & \mbox{for} \; |q| \leq q_c \\
0 &    \mbox{for} \; |q| >  q_c \end{array} \right. \; , 
\label{boxpot} \\
v_{\rm gauss} (q) & = & v \, e^{-(q/q_c)^2/2}  \; ,
\label{gausspot} \\
v_{\rm exp} (q) & = & v \, e^{-|q|/q_c} \; ,
\label{exppot} \\
v_{\rm quart}(q) & = & \frac{v}{1+  \left(\frac{q}{q_{\rm c}} \right)^4 } \; .
\label{quartpot}  
\end{eqnarray}
All potentials have the same $q=0$ value $v(0)=v$ and thus the 
same LL parameter $K$ and renormalized Fermi velocity $\hat v_{\rm F}$ 
(see Eqs.~(\ref{LLpardef}) and (\ref{renvfdef})).
In equilibrium they give the same low-energy LL physics.  
To obtain this also for the
potential Eq.~(\ref{pfusch}) one has to choose 
\begin{eqnarray}
g= \frac{1}{2} \, \frac{\hat v}{\sqrt{1+\hat v}} \; .
\label{gdef}
\end{eqnarray}
In Sec.~\ref{timeevolution} we show that the quench dynamics has aspects 
which are equal for different choices of the potential while other features 
depend on the potential. Before 
discussing this, in the next section we compute $G^{\rm st}(x)$. 
The steady-state Green function and thus $n^{\rm st}(k)$ turns out to be universal 
in the sense of the LL phenomenology.
We furthermore compare  $n^{\rm st}(k)$ to the corresponding equilibrium 
momentum distribution function computed for the
same interaction strength. 

\section{Steady-state and equilibrium expectation values} 
\label{ssee}

\subsection{Infinite-time limit and equilibrium}
\label{ssee1}

The infinite-time 
steady-state value of $G_t(x)$ Eqs.~(\ref{GTL}) and (\ref{GTLWW1WW2})
can be obtained straightforwardly. We first consider  the 
quench out of the noninteracting ground state. 
For fixed large $t$ the cosine term 
in Eq.~(\ref{GTL}) with argument linear in $t$ 
becomes a rapidly oscillating function of $k$ and averages out. In the limit $t \to \infty$ 
we thus end up with 
\begin{eqnarray}
 \mbox{} \hspace{-1cm} G^{\rm st}(x) =  \lim_{t \to \infty} G_t(x) = 
\frac{i}{2 \pi} \, 
\frac{e^{-i k_{\rm F} x}}{x +i0} 
\exp{ \left\{ \int_0^\infty  \!\!\! dk \, \frac{4 s^2(k) c^2(k)}{k} \left( \cos{[k x]} -1
    \right)  \right\} } \; . \label{Gst}
\end{eqnarray} 

The fermionic momentum distribution function in the steady state $n^{\rm st}(k)$, 
that is the Fourier transform of Eq.~(\ref{Gst}), can {\em close} to $k_{\rm F}$ be 
computed {\em without} any specific assumptions for the two-particle potential
using asymptotic analysis  \cite{Orszag99}. One can closely follow the steps 
of  Ref.~\cite{Meden99} for the equilibrium ground-state momentum distribution
function obtained from Fourier transforming Eq.~(\ref{GTLequ}). Next we briefly outline
those. The leading large $|x|$ behavior of Eq.~(\ref{GTLequ}) 
is given by the integrand at small momenta and dominates the Fourier transform close 
to $k_{\rm F}$ leading to
\begin{eqnarray}
|n^{\rm gs}(k) -1/2| \sim \left| \Delta \hat k \right|^{\gamma_{\rm gs}}
\label{nkLLgs}
\end{eqnarray}   
independent of the details of the potential \cite{Meden99}. 
Here we have introduced the relative dimensionless momentum 
\begin{eqnarray}
\Delta \hat k = \frac{k-k_{\rm F}}{q_{\rm c}} \; .
\end{eqnarray}  
The  {\em equilibrium anomalous dimension} $\gamma_{\rm gs}$
reads (we drop the index i)
\begin{eqnarray}
\gamma_{\rm gs} = 2 s^2(0) = \frac{1}{2} \left( K  + K^{-1} -2 \right)
\label{anomalousgs}
\end{eqnarray} 
Equation (\ref{nkLLgs}) holds for all two-particle potentials 
within the TL model as long as $0 < \gamma_{\rm gs} < 1$, that is
for sufficiently small interactions. It implies 
a power-law singularity of the first derivative 
of $n^{\rm gs}(k)$ at $k_{\rm F}$. For larger interactions 
$n^{\rm gs}(k)$ goes {\em linearly} through $k_{\rm F}$ and 
singularities appear in higher order derivatives \cite{Meden99}. 

In complete analogy one obtains for the steady-state momentum distribution 
function $n^{\rm st}(k)$ after a quench out of the noninteracting state
\begin{eqnarray}
|n^{\rm st}(k) -1/2| \sim \left|\Delta \hat k \right|^{\gamma_{\rm st}}
\label{nkLLnonequ}
\end{eqnarray}   
with the {\em nonequilibrium anomalous dimension}
\begin{eqnarray}
\gamma_{\rm st} & = & 4 s^2(0) c^2(0)  = \frac{1}{4} \, \frac{\hat v^2(0)}{1+ \hat v(0)}  \nonumber \\
 & = & \frac{1}{4} \, \left( K^2 + K^{-2} -2 \right) \; .
\label{anomalousst}
\end{eqnarray} 
For Eq.~(\ref{nkLLnonequ}) describing the leading behavior close to $k_{\rm F}$ the same 
restrictions as outlined in connection with Eq.~(\ref{nkLLgs}) hold (with $\gamma_{\rm gs}$ 
replaced by $\gamma_{\rm st}$). 
In the specific case of the potential Eq.~(\ref{pfusch}) the results 
Eqs.~(\ref{nkLLnonequ}) and (\ref{anomalousst}) were obtained in Ref.~\cite{Cazalilla06}. 
The ad hoc replacement  Eq.~(\ref{pfusch_2}) is not required here since 
in the limit $t \to \infty$, $\omega(k)$ drops out. 
Interestingly, $\gamma_{\rm st}$ in the steady state of the TL model 
differs from the exponent $\gamma_{\rm gs}$ found in the ground state of $H_{\rm TL}$ 
at the same interaction strength. In the next subsection we return to this issue. 

Quenching between two repulsive interactions of arbitrary strength the steady-state Green 
function follows from Eq.~(\ref{GTLWW1WW2}) by dropping the cosine term with time 
dependent argument
 \begin{eqnarray}
\mbox{} \hspace{-2cm} && G^{\rm st}(x) =   \frac{i}{2 \pi} \, 
\frac{e^{-i k_{\rm F} x}}{x +i0} \exp  \left\{ 2 \int_0^\infty  \!\!\! dk \, \frac{\cos{[k x]} -1 }{k} 
\right. \nonumber \\ 
\mbox{} \hspace{-2cm} && \hspace{1cm}
\times \left. \left( c_{\rm f}^2(k) \left[ c_{\rm f}(k) s_{\rm i}(k) - s_{\rm f}(k) c_{\rm i}(k)
\right]^2 + s_{\rm f}^2(k) \left[ c_{\rm f}(k) c_{\rm i}(k) - s_{\rm f}(k) s_{\rm i}(k)
\right]^2 \right) \phantom{\frac{1}{2}} \!\!\!\! \!\! \right\}  \; .
\label{GTLWW1WW2st}
\end{eqnarray}
Independently of the details of the momentum dependence of the initial and final potential Fourier 
transforming again leads to Eq.~(\ref{nkLLnonequ}) with the nonequilibrium anomalous dimension given by 
\begin{eqnarray}
\mbox{} \hspace{-1cm} \gamma_{\rm st}  =  2  \left( c_{\rm f}^2(0) \left[ c_{\rm f}(0) s_{\rm i}(0) - s_{\rm f}(0) c_{\rm i}(0)
\right]^2 + s_{\rm f}^2(0) \left[ c_{\rm f}(0) c_{\rm i}(0) - s_{\rm f}(0) s_{\rm i}(0)
\right]^2 \right) \; .
\label{anomalousstgeneral}
\end{eqnarray} 
Using Eq.~(\ref{scK}) (supplemented by the indices i and f) $\gamma_{\rm st}$ can be written as a function 
of the LL parameters $K_{\rm i}$ and $K_{\rm f}$ associated to the two interactions. As this does not 
have a simple form we refrain from presenting it here.  

Power-law behavior of correlation functions is a typical feature of LL physics. 
Within the TL model we thus find universal behavior known from equilibrium: 
the steady-state momentum distribution function close to $k_{\rm F}$ is independently of 
the details of the momentum dependence of the potential characterized 
by a power law with an exponent which can be expressed in terms of $K$ or $K_{\rm i}$ and 
$K_{\rm f}$. It would be 
very interesting to investigate if this universality of the steady-state expectation value 
{\em extends beyond the TL model.} For this it would be necessary to analytically or numerically 
compute $n^{\rm st}(k)$ for other models from the LL universality class, 
e.g.~lattice models such as the model of spinless fermions with nearest-neighbor 
hopping and interaction, for which $K$ is known from other considerations 
(e.g.~the Bethe ansatz or numerics) 
\cite{Voit95,Giamarchi03,Schoenhammer05}.\footnote{One crucial difference of a generic lattice model 
to the TL model is the nonlinearity of the single-particle dispersion of the fermions. Similar to the 
momentum dependence of the two-particle potential this might have an effect on the transient dynamics. 
For a review on the effect of the nonlinear dispersion on equilibrium LL physics, see Ref.~\cite{Imambekov11}.} 
One could then 
extract the possible power-law exponent (as a function of $|k-k_{\rm F}|$) for fixed model 
parameters and compare to the expression in the second line of Eq.~(\ref{anomalousst})
or Eq.~(\ref{anomalousstgeneral}) (the latter supplemented by Eq.(\ref{scK})).  

In Refs.~\cite{Cazalilla06} and \cite{Iucci09} it was shown that $G^{\rm st}(x)$ 
(and also the stationary small momentum density correlation function) of the TL model 
can be computed as an average with a nonthermal statistical operator of a generalized 
Gibbs ensemble (GGE) with the eigenmode occupancies as the underlying 
set of integrals of motion. In the context of quenches the concept of 
GGEs was introduced to describe the (possible) stationary-state value of time 
evolved observables in systems with {\em many} conserved quantities 
\cite{Rigol07,Cramer08,Barthel08}. We verified that a similar GGE can be used if 
(i) the full momentum dependence of the potential is kept and (ii) for 
quenches between two interactions of arbitrary strength.

\subsection{A factor of two}

In systems which after a quench are expected to evolve into a thermal stationary 
state (described by a canonical ensemble), on intermediate time scales the appearance 
of  pre-thermalized quasi 
stationary state was observed when computing the time evolution out of the noninteracting 
ground state in the {\em weak coupling limit} \cite{Moeckel08,Moeckel09}. The quasi 
stationary state is characterized by 
observables oscillating for some time interval around a constant value 
which is different from the stationary one, the latter being reached for much 
larger times. Averaging observables like 
the fermionic momentum distribution function (at fixed $k$) 
over times in which
the system is stuck in the pre-thermalized state gives values which agree to 
the corresponding ground state expectation values of the interacting system 
described by $H_{\rm f}$ up to a characteristic factor of 
two \cite{Moeckel08,Moeckel09}. It was later argued \cite{Kollar11} that the pre-thermalized 
states correspond to the nonthermal steady state of systems with a sufficiently large  number
of  integrals of motion. We thus expect to find these factors of two for the TL model, which falls 
into this class of systems, 
when considering the weak coupling limit. To keep this section compact we here exclusively
consider the case of the quench out of the noninteracting ground state.

Following Refs.~\cite{Moeckel08,Moeckel09} for the fermionic momentum distribution 
function we expect
\begin{eqnarray}
2 \left[  n_{\rm gs}(k)-  n^0_{\rm gs}(k) \right] = n^{\rm st} (k)-  n^0_{\rm gs}(k)
\label{nexpect}
\end{eqnarray}
to hold up to order $\hat v^2$. Here 
\begin{eqnarray}
n^0_{\rm gs}=\Theta(k_{\rm F}-k)
\end{eqnarray}
is the noninteracting ground-state momentum distribution function (of right movers). 
Using the expansion (see Eq.~(\ref{manydefs})) 
\begin{eqnarray}
s^2(k) = \frac{1}{16} \, \hat v^2(k)  
+ {\mathcal O}\left( \hat v^3(k)\right)
\label{expfund}
\end{eqnarray}
in Eqs.~(\ref{GTLequ}) and (\ref{Gst}) we obtain
\begin{eqnarray} 
2 G^{\rm gs}(x)
- G^{\rm st}(x) = \frac{i}{2 \pi} \, \frac{e^{-i k_{\rm F} x}}{x+ i0} +    
{\mathcal O}\left( \hat v^4 \right) \; .
\end{eqnarray}
The equilibrium ground-state Green function $G^{\rm gs}(x)$ Eq.~(\ref{GTLequ}) 
is computed with the interaction after the quench that is with 
$s^2_{\rm i}(k) \to s^2(k)$.  
Fourier transformation then gives Eq.~(\ref{nexpect}) up to {\em third order.}

We take the opportunity and also compare the anomalous dimensions of the steady state 
and the ground state characterizing the momentum distribution function close to $k_{\rm F}$. 
The former is given in Eq.~(\ref{anomalousst}), the latter in  Eq.~(\ref{anomalousgs}).
Expanding in the interaction using Eq.~(\ref{expfund}) gives 
\begin{eqnarray}
2 \gamma_{\rm gs} - \gamma_{\rm st} = {\mathcal O}\left( \hat v^4 \right) \; .
\end{eqnarray}
Interestingly, we thus also obtain the relative factor of two in the power-law exponents of the momentum 
distribution functions at sufficiently weak coupling.

Our comparison of the steady-state and ground-state momentum distribution function 
of the TL model provides additional evidence that the notion of a pre-thermalized state 
in a general model \cite{Moeckel08,Moeckel09} and its relation to the steady state in a 
model with many integrals of motion \cite{Kollar11} is indeed meaningful.

\section{Time evolution after a quench} 
\label{timeevolution}

We now evaluate the momentum integral of Eq.~(\ref{GTL}) for the Green function 
using the different two-particle potentials introduced at the end of Sec.~\ref{modbos} 
as well as with the ad hoc replacement Eq.~(\ref{pfusch_2}) (and the corresponding 
Eq.~(\ref{pfusch})). In addition, we Fourier transform the Green function 
(see Eq.~(\ref{nofk})). This way we obtain explicit results for the time evolution 
of the fermionic momentum distribution function after an interaction quench out of the noninteracting
ground state. The results obtained for the different potentials are compared.  

\subsection{Analytical insights}

\subsubsection{Jump at $k_{\rm F}$} 

We start out with an analytical result which can be obtained independent of the  
form of the two-particle potential (and even for the ad hoc procedure). 
Using Eq.~(\ref{GTL}) it is straightforward to 
show that the jump $Z(t)$ of $n(k,t)$ at $k_{\rm F}$  defined as 
\begin{eqnarray} 
Z(t) = \lim_{k \nearrow k_{\rm F}} n(k,t) -  \lim_{k \searrow k_{\rm F}} n(k,t)
\end{eqnarray}
is given by 
\begin{eqnarray}
Z(t) = \exp{ \left\{ - \int_0^\infty  \!\!\! dk \, \frac{4 s^2(k) c^2(k)}{k} 
\left(  1- \cos{\left[2 \omega(k) t\right]} \right)  \right\} } \; .
\end{eqnarray}
For large times the remaining momentum integral can be performed using asymptotic analysis 
\cite{Orszag99} leading to the result
\begin{eqnarray} 
Z(t) \sim \hat t^{- \gamma_{\rm st}} 
\end{eqnarray}
first obtained for the ad hoc procedure in Ref.~\cite{Cazalilla06}. We here introduced the dimensionless
time
\begin{eqnarray} 
\hat t = v_{\rm F} q_{\rm c} \, t \; . 
\end{eqnarray}

\subsubsection{The ad hoc procedure}

With the assumption Eq.~(\ref{pfusch}) and  the replacement Eq.~(\ref{pfusch_2}) the momentum integral 
in Eq.~(\ref{GTL}) can be performed analytically leading to 
\begin{eqnarray}
\mbox{} \hspace{-2.5cm} G^{\rm adhoc}(x,t) = \frac{i}{2\pi} \, l_{\rm c}^{\gamma_{\rm st}} \, e^{-i k_{\rm F} x} 
\, \frac{\left[ \left(x+2 \tilde v_{\rm F} t \right)^2 + l_{\rm c}^2 \right]^{\gamma_{\rm st}/4} 
\left[ \left(x - 2 \tilde v_{\rm F} t \right)^2 + l_{\rm c}^2 \right]^{\gamma_{\rm st}/4} }{
\left[x +i0 \right] \left[ x^2 + l_{\rm c}^2 \right]^{\gamma_{\rm st}/2}  \left[ \left(2 \tilde v_{\rm F} t \right)^2 
+ l_{\rm c}^2 \right]^{\gamma_{\rm st}/2}} \; , \;\; l_{\rm c} = q^{-1}_{\rm c} .
\end{eqnarray}
Comparing Eq.~(\ref{gdef}) and the first line of Eq.~(\ref{anomalousst}) we find that 
$g^2=\gamma_{\rm st}$ and have accordingly replaced the former by the latter. 
Using the above expression the momentum distribution function can be written as
\begin{eqnarray}
\mbox{} \hspace{-2.5cm} && \Delta n^{\rm adhoc}(k,t) = n^{\rm adhoc}(k,t) - \lim_{t \to \infty} n^{\rm adhoc}(k,t) \nonumber \\
\mbox{} \hspace{-2.5cm}  && = \frac{i l_{\rm c}^{\gamma_{\rm st}} }{2\pi} \int_{-\infty}^\infty dx \, 
\frac{e^{i(k-k_{\rm F} )x} }{\left[x +i0 \right] \left[ x^2 + l_{\rm c}^2 \right]^{\gamma_{\rm st}/2} }
\left\{    \frac{\left[ \left(x+2 \tilde v_{\rm F} t \right)^2 + l_{\rm c}^2 \right]^{\gamma_{\rm st}/4} 
\left[ \left(x - 2 \tilde v_{\rm F} t \right)^2 + l_{\rm c}^2 \right]^{\gamma_{\rm st}/4} }{
\left[ \left(2 \tilde v_{\rm F} t \right)^2 
+ l_{\rm c}^2 \right]^{\gamma_{\rm st}/2}}
-1
\right\} \nonumber \\
\mbox{} \hspace{-2.5cm}  && = -
\frac{1}{\pi} \, \frac{1}{(2 \hat v_{\rm F} \hat t)^{\gamma_{\rm st}}} \, \int_0^\infty dy \, 
\frac{\sin\left[ 2  \hat v_{\rm F} \Delta \hat k \hat t y \right]}{y  
\left[ y^2 + (2  \hat v_{\rm F} \hat t)^{-2} \right]^{\gamma_{\rm st}/2}}  
\nonumber \\ 
\mbox{} \hspace{-2.5cm} &&   \mbox{} \hspace{3.cm}  \times \left\{  
\frac{\left[ (1+y)^2 +  (2  \hat v_{\rm F} \hat t)^{-2} \right]^{\gamma_{\rm st}/4} \left[ (1-y)^2 +  (2  \hat v_{\rm F} \hat t)^{-2} 
\right]^{\gamma_{\rm st}/4} }
{\left[ 1+  (2  \hat v_{\rm F} \hat t)^{-2} \right]^{\gamma_{\rm st}/2}} 
-1
\right\} \; .
\label{adhocana}
\end{eqnarray} 
For $\gamma_{\rm st} < 2$ the integrand excluding the sine factor has a cusp at 
$y=1$---for $\hat t \to \infty$ the first derivative for $y \nearrow 1$ goes to 
minus infinity while for  $y \searrow 1$ it goes to plus infinity. 
For  fixed $k \neq k_{\rm F}$ (for the behavior at $k=k_{\rm F}$, see 
above) this cusp leads to a nonanalytic power-law decay of 
$\Delta n^{\rm adhoc}(k,t)$ given by  
\begin{eqnarray}
\Delta n^{\rm ad hoc}(k,t) \sim 
\hat t^{-1-\frac{3}{2} \gamma_{\rm st}} \; \sin\left[ 2  \hat v_{\rm F} \Delta \hat k \hat t\right] \; ,
\label{nklargetadhoc}
\end{eqnarray}
which follows from restricting the integral to a small region around $y=1$. The integral also has 
regular parts starting with a term $\sim \hat t^{-2}$. Therefore Eq.~(\ref{nklargetadhoc})
gives the asymptotic behavior as long as $\frac{3}{2} \gamma_{\rm st} < 1$. Although 
it is possible to analyze $\Delta n^{\rm adhoc}(k,t)$ for stronger interactions we here refrain
from doing so as we are mainly interested in the limit of weak to intermediate interactions.          
Loosely speaking for large $\hat t$ the smooth 
contributions away from the cusp average out due to the oscillatory term while close to the 
cusp the integrand changes quickly providing the leading contribution.  
This argument implies that the (dimensionless) time scale $\hat t_{\rm p}$ 
at which the power-law decay sets in increases for decreasing $|k - k_{\rm F}|$
because the frequency of the $y$-oscillation decreases: $\hat t_{\rm p} \sim 
\left| \Delta \hat k \right|^{-1}$.
The closer the momentum is to the nonanalyticity at $k_{\rm F}$ the larger the time scale 
on which the asymptotic behavior sets in. In summary, within the ad hoc procedure 
the momentum distribution function at fixed $k$ and for $\hat t \gg \hat t_{\rm p}$ 
approaches its stationary value in an oscillatory fashion with (dimensionless) frequency  
$2  \hat v_{\rm F} \Delta \hat k $ and an 
amplitude decaying as a power-law in $\hat t$ with exponent $1+\frac{3}{2} \gamma_{\rm st}$ 
(as long as the interaction is not too strong, that is for $\frac{3}{2} \gamma_{\rm st}  <1$).

\begin{figure}[t]
\begin{center}
   \includegraphics[width=.55\linewidth,clip]{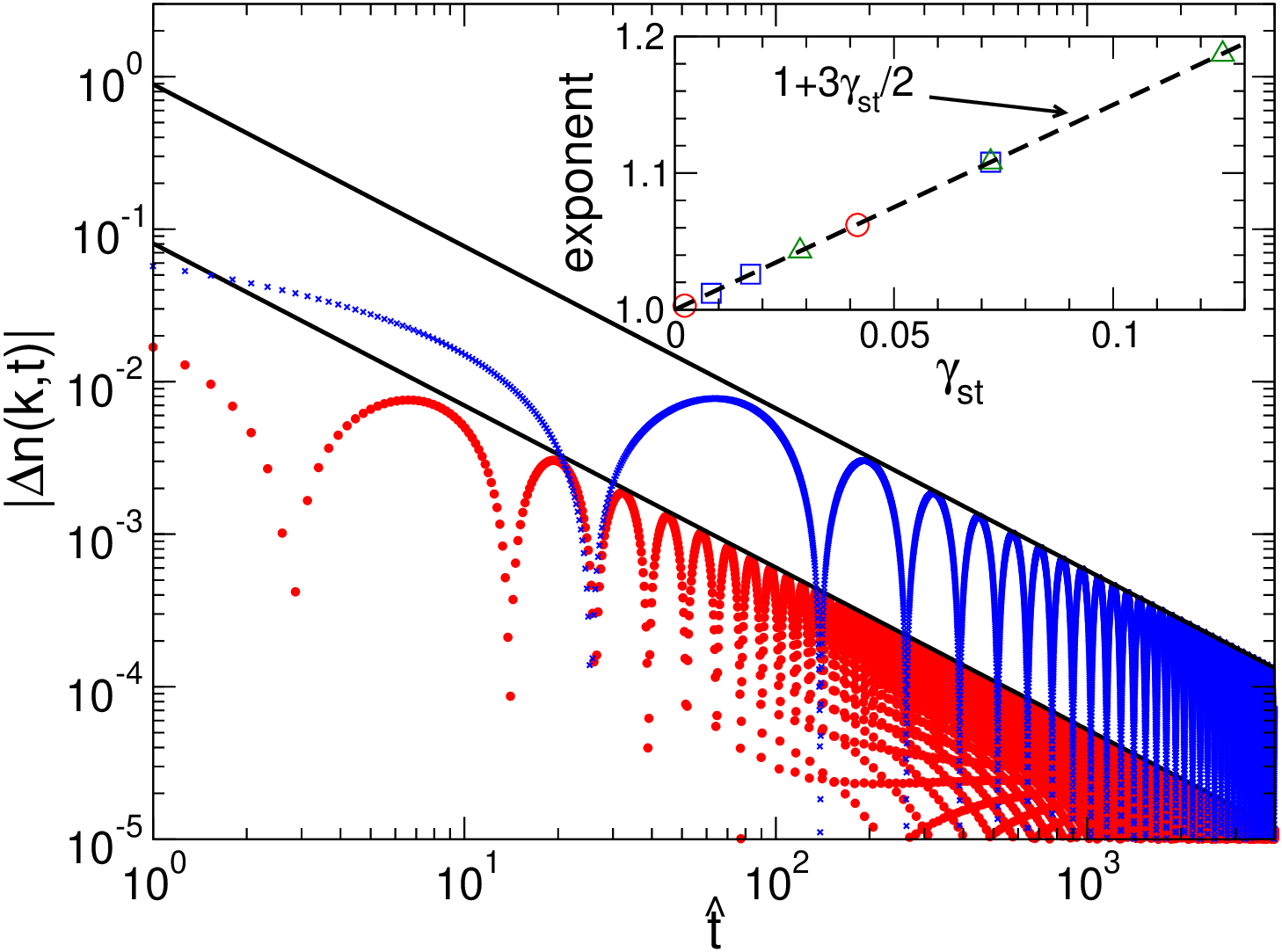}
   \caption{The time dependence of $ | \Delta n (k,t)|$ for the ad hoc regularization 
     with $\hat v=0.5$ at fixed $\Delta \hat k =0.1$ (red circles)  and  
     $\Delta \hat k =0.01$ (blue crosses) on a double logarithmic scale. The solid lines show 
     power-law fits to the envelope. Inset:   
     Exponents extracted for different relative momenta  $\Delta \hat k$ (circles: 0.8; 
     squares: 0.4; triangles: 0.1) as a function of $\gamma_{\rm st}$, that is the 
     interaction strength. The dashed line is the analytical result $1+3 \gamma_{\rm st}/2$.}
   \label{fig1}
\end{center}
\end{figure}

These analytical insights can be confirmed numerically by performing the integral in Eq.~(\ref{adhocana}). 
Figure \ref{fig1} shows $\left| \Delta n^{\rm ad hoc}(k,t) \right|$ for 
fixed  $\Delta \hat k=0.1$ (red circles) as well as  $\Delta \hat k=0.01$ (blue crosses) and $\hat v=0.5$  
on  a double-logarithmic scale. The solid lines are power-law fits (for times 
$\hat t \in [400,4000] $) to the envelope. 
In the inset exponents 
extracted along this line for different $\hat v$ (and thus $\gamma_{\rm st}$; see Eq.~(\ref{anomalousst})) 
and $\Delta \hat k$ are presented. They nicely fall onto the analytical prediction 
$1+\frac{3}{2} \gamma_{\rm st} $ shown as the dashed line.   Consistent with the above analytical 
result the asymptotic behavior is reached faster the larger $\Delta \hat k$ (compare the 
data for  $\Delta \hat k=0.1$ and $0.01$ in the main panel of Fig.~\ref{fig1}).

\subsubsection{The box potential}

Analytical progress is also possible in the case of the box potential Eq.~(\ref{boxpot}). 
Then the argument of the exponential function in Eq.~(\ref{GTL}) simplifies to 
\begin{eqnarray}
I & = &  \int_0^\infty  dk \, \frac{4 s^2(k) c^2(k)}{k} \left( \cos{[k x]} -1
    \right) \left(  1- \cos{\left[2 \omega(k) t\right]} \right)  \nonumber \\
& = &  \gamma_{\rm st} \int_0^1  dk \, \frac{ \cos{[k q_{\rm c} x]} -1}{k} 
\left(  1- \cos{\left[2 \hat v_{\rm F} k \hat t \right]} \right)  \; .
\end{eqnarray}
For asymptotically large $\hat t$ the integral can be evaluated using integration by parts  
\begin{eqnarray}
\mbox{} \hspace{-1cm} I = \gamma_{\rm st} \int_0^1  dk \, \frac{ \cos{[k q_{\rm c} x]} -1}{k} 
- \frac{\gamma_{\rm st} }{2 \hat v_{\rm F} \hat t} \left( \cos{[q_{\rm c} x]} -1 \right) \sin{(2 \hat v_{\rm F} \hat t)}
+ {\mathcal O}(\hat t^{-2}) \; .
\end{eqnarray}
From this it straightforwardly follows that 
\begin{eqnarray}
\Delta n^{\rm box}(k,t) \sim \frac{\sin\left( 2  \hat v_{\rm F} \hat t \right)}{\hat t} \; .
\label{boxasym}
\end{eqnarray} 
The two characteristic differences to the asymptotic dynamics of the ad hoc procedure 
Eq.~(\ref{nklargetadhoc}) namely the (i) independence of the decay 
exponent from the interaction strength and the (ii) independence of the oscillation frequency from 
the momentum $k-k_{\rm F}$ at which the distribution function is evaluated can be traced back to 
the nonanalyticity of the box potential; the long-time dynamics is dominated by the position 
of the jump in the potential, that is the upper boundary of the momentum integral in $I$.

\begin{figure}[t]
\begin{center}
   \includegraphics[width=.55\linewidth,clip]{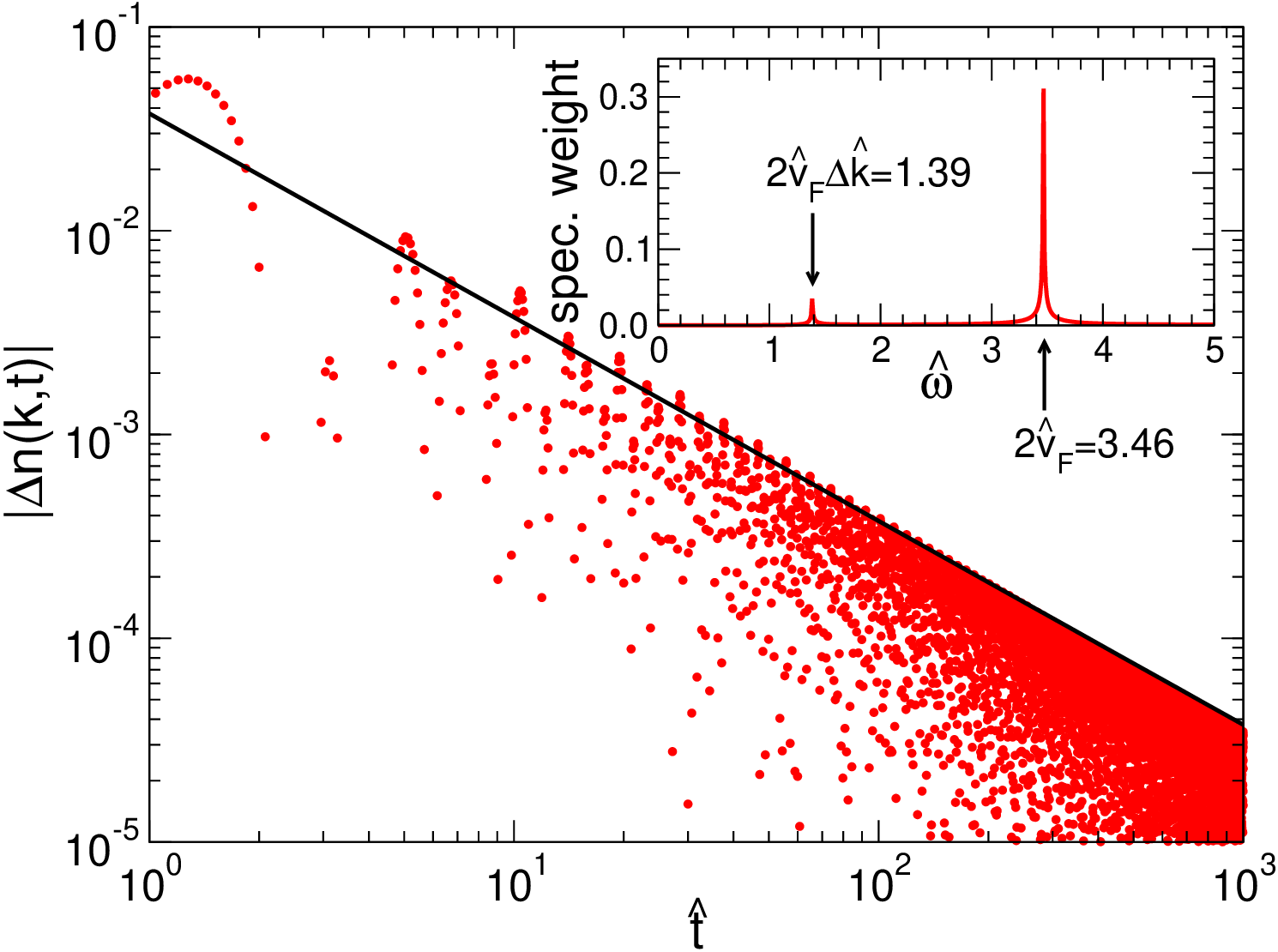}
   \caption{The time dependence of $ | \Delta n (k,t)|$ for the box potential
     with $\hat v=2$ at fixed $\Delta \hat k =0.4$ (red circles). The solid 
     line shows a power law $\sim \hat t^{-1}$ which fits the data for sufficiently large 
     $\hat t \gtrapprox 200$. Inset: Fourier spectrum of the data from the main plot.}
   \label{fig2}
\end{center}
\end{figure}

Figure \ref{fig2} shows $\left| \Delta n^{\rm box}(k,t) \right|$ for  fixed 
$\Delta \hat k=0.4$ and $\hat v=2$  
obtained by numerically performing the momentum integral in Eq.~(\ref{GTL}) and the Fourier integral 
(with respect to position) Eq.~(\ref{nofk}). In the data a second 
smaller frequency than  $2  \hat v_{\rm F}$ is observable. 
A Fourier analysis (with respect to time) shows that it is equal to the frequency 
found for $\Delta n^{\rm ad hoc}(k,t)$ and given by   
$2  \hat v_{\rm F} \Delta \hat k$; see the inset of Fig.~\ref{fig2}. 
In fact, the numerical data are consistent with the long-time dynamics for the box potential being 
dominated by the interplay of the term Eq.~(\ref{boxasym}) and another one of the form 
Eq.~(\ref{nklargetadhoc}). Considering a fixed time interval the $\hat t^{-1}$ decay 
is only clearly observable if the interaction and thus $\gamma_{\rm st}$ is not too small and 
the term Eq.~(\ref{nklargetadhoc}) can be neglected compared to  Eq.~(\ref{boxasym}). Consistently, 
for such interactions the low-energy peak in the frequency spectrum carries a much lower weight than 
the high-energy one; see the inset of Fig.~\ref{fig2}.  

It is possible to provide analytical evidence for the appearance of the second term 
Eq.~(\ref{nklargetadhoc}) at large $\hat t$. 
To this end one uses that the argument $I$ of the exponential function in Eq.~(\ref{GTL}) can be 
further evaluated  
\begin{eqnarray}
I & = &  \gamma_{\rm st} \left[ \mbox{Ci}(|q_{\rm c}x|) - \gamma - \ln ( |q_{\rm c}x| )\right]
+  \gamma_{\rm st} \left[ \mbox{Ci}(2 \hat v_{\rm F} \hat t) - \gamma - \ln (2 \hat v_{\rm F} \hat t )  \right]
\nonumber \\*
&& -  \frac{\gamma_{\rm st}}{2} \left[ \mbox{Ci}(|q_{\rm c}x + 2 \hat v_{\rm F} \hat t|) 
- \gamma - \ln ( |q_{\rm c}x +2 \hat v_{\rm F} \hat t| )\right]\nonumber \\*
&&
-  \frac{\gamma_{\rm st}}{2} \left[ \mbox{Ci}(|q_{\rm c}x - 2 \hat v_{\rm F} \hat t|) 
- \gamma - \ln ( |q_{\rm c}x -2 \hat v_{\rm F} \hat t| )\right]
\end{eqnarray}
with the integral cosine function $\mbox{Ci}$ and the Euler constant $\gamma$. With this 
 $\left| \Delta n^{\rm box}(k,t) \right|$ can be brought into a form similar to Eq.~(\ref{adhocana})
\begin{eqnarray}
\mbox{} \hspace{-2.5cm}  \Delta n^{\rm box}(k,t)  & = & -  \frac{1}{\pi}
\frac{\exp{(-\gamma \gamma_{\rm st})}}{(2 \hat v_{\rm F} \hat t)^{\gamma_{\rm st}}} \, \int_0^\infty dy \, 
\frac{\sin\left[ 2  \hat v_{\rm F} \Delta \hat k \hat t y \right]}{y^{1+\gamma_{\rm st}}}  
\Big[  (1+y)^{\gamma_{\rm st}/2} |1-y|^{\gamma_{\rm st}/2}  
\nonumber \\ 
\mbox{} \hspace{-2.5cm} &&   \times \exp{ \left\{ \gamma_{\rm st} \left[ \mbox{Ci}(2 \hat v_{\rm F} \hat t) - 
 \mbox{Ci}(2 \hat v_{\rm F} \hat t[y+1])/2 - 
 \mbox{Ci}(2 \hat v_{\rm F} \hat t|y-1|)/2    \right] \right\} } 
-1
\Big] .
\label{boxana}
\end{eqnarray}   
Leaving out the sine factor and considering large $\hat t$  the integrand has similar to the one 
of Eq.~(\ref{adhocana}) a cusp at $y=1$. This again leads to a nonanalytic term of the form 
Eq.~(\ref{nklargetadhoc}). In fact, the integrands 
(as a function of $y$) of the box potential and the ad hoc procedure 
coincide close to $y=1$ up to the crucial difference that for the box potential the cusp-like behavior is 
modulated by an oscillation with frequency $ 2  \hat v_{\rm F} \hat t$ (associated to the oscillatory 
behavior of $\mbox{Ci}$). Taking everything together the  appearance of the two terms  
Eqs.~(\ref{nklargetadhoc}) and (\ref{boxasym}) is thus plausible from the analytics. 

To observe the power-law decay of $\left| \Delta n^{\rm box}(k,t) \right|$ for fixed $k-k_{\rm F}$ 
in a given time interval (at large times) one again has to stay away from the nonanalyticity 
(the jump for finite $t$) at $k=k_{\rm F}$: the smaller $\left| \Delta \hat k \right|$ the longer 
it takes before the asymptotic power-law behavior sets in. 

\subsubsection{Stationary points in $\omega(k)$}

For the regular potentials  Eqs.~(\ref{gausspot})-(\ref{quartpot}) the dispersion 
$\omega(k)$ Eq.~(\ref{manydefs}) is a nonlinear function of the momentum. For each 
of the potentials a critical interaction strength $\hat v_{\rm c}$ exists beyond which 
$\omega(k)$ has two stationary points. For the Gaussian potential one e.g. finds 
$\hat v_{\rm c}=e^2 \approx 7.39$. For $\hat v > \hat v_{\rm c}$ the large time dynamics 
of Eq.~(\ref{GTL}) is dominated by these stationary points. Using the stationary phase method 
\cite{Orszag99} it is straightforward to show that on asymptotic time scales 
\begin{eqnarray}
\mbox{} \hspace{-0.5cm}  \Delta n(k,t) 
\sim c_1 \, \frac{\sin{[2 \hat \omega(k_1) \hat t+\phi_1]}}{\sqrt{ \hat t}} +  
 c_2 \, \frac{\sin{[2 \hat \omega(k_2) \hat t+\phi_2]}}{\sqrt{\hat t}} \; ,
\label{nktstatpoint}
\end{eqnarray}   
with amplitudes $c_{1/2}$, phases $\phi_{1/2}$, the two stationary points 
$k_{1/2}$, and the dimensionless dispersion $\hat \omega(k) = 
\omega(k)/(v_{\rm F} q_{\rm c})$.\footnote{For an example in which this type of behavior dominates for all 
interaction strengths, see Ref.~\cite{Kennes10}.}  
Here we are primarily interested in the behavior 
at small to intermediate interactions (see also the above subsection on the ad hoc procedure) and thus 
focus on $\hat v < \hat v_{\rm c}$ from now on.

\subsection{Numerical results}

For the Gaussian Eq.~(\ref{gausspot}), the exponential Eq.~(\ref{exppot}), and the quartic 
potential  Eq.~(\ref{quartpot}) 
we did not succeed in obtaining analytical results for $\Delta n(k,t)$. The following analysis 
thus solely relies on the numerical evaluation of the momentum integral in Eq.~(\ref{GTL}) 
and the successive Fourier integral Eq.~(\ref{nofk}). As the integrands are oscillatory functions
one has to use routines which are adopted to this situation. Furthermore, for large times 
$ | \Delta n (k,t)|$ becomes very small (of order $10^{-8}$ and smaller) which requires a very 
accurate evaluation of the integrals. This limits the numerics and reliable results can 
be obtained up to times of the order of $\hat t = 5000$.  
  
\begin{figure}[t]
\begin{center}
   \includegraphics[width=.55\linewidth,clip]{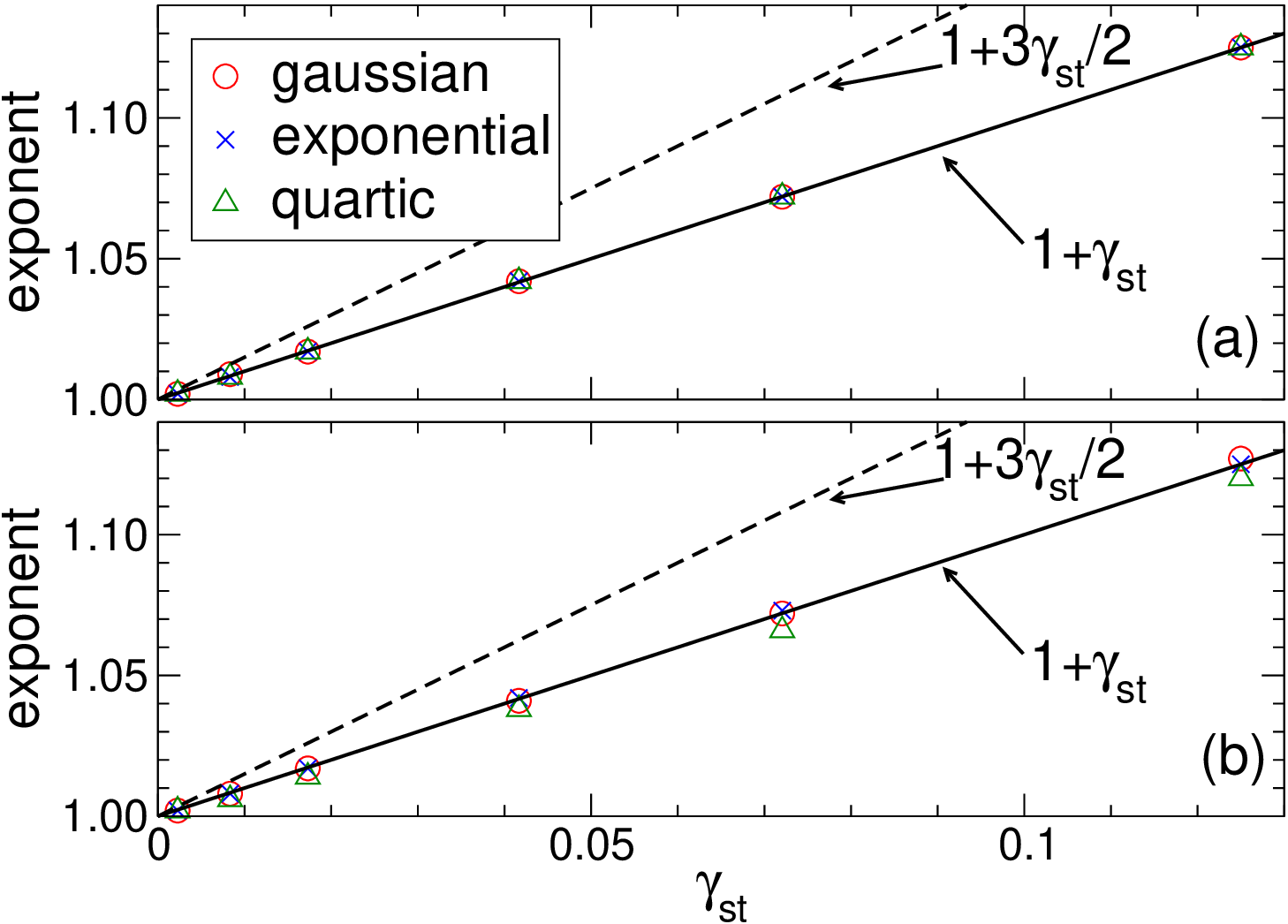}
   \caption{Exponents of the power-law decay of  $ | \Delta n (k,t)|$ for different potentials 
     (symbols) as a function of  $\gamma_{\rm st}$, that is the interaction strength (see Eq.(\ref{anomalousst})). 
     The solid line shows $1+\gamma_{\rm st}$, the dashed one the exponent $1+3 \gamma_{\rm st}/2$ obtained 
     within the ad hoc procedure. (a) Fixed momentum  $\Delta \hat k=0.8$ and 
     (b) $\Delta \hat k=0.4$.}
   \label{fig3}
\end{center}
\end{figure}

From the above analytical consideration we expect that (i) $ | \Delta n (k,t)|$ decays 
as a power law in $\hat t$ and that (ii) this should be observable on moderately large 
times if $\left| \Delta \hat k \right|$ does not become too small. Our numerical results 
for $ | \Delta n (k,t)|$ (not shown; the general form is similar to the data of Fig.~\ref{fig1}) 
are consistent with this expectation. In Fig.~\ref{fig3} we show how 
exponents extracted by a power-law fit of $ | \Delta n (k,t)|$  for times 
$\hat t \in[100,2400]$ depend on  $\gamma_{\rm st}$, that is  $\hat v$ (symbols).
In Fig.~\ref{fig3}(a) the momentum is fixed at $\Delta \hat k=0.8$ 
and in Fig.~\ref{fig3}(b) at $\Delta \hat k=0.4$. The dashed line shows the result 
$1+3 \gamma_{\rm st}/2$ obtained for the exponent within the ad hoc procedure. Obviously the 
data for the three potentials coincide and differ from $1+3 \gamma_{\rm st}/2$ as well as  
the exponent 1 found for the box potential. In Fig.~\ref{fig3}(a) they instead nicely 
fall onto the line $1+ \gamma_{\rm st}$. For 
smaller $\Delta \hat k=0.4$ the data slightly scatter around this line but are still consistent
with it. The largest deviations are observed for the quartic 
potential (see below). Already at this stage of the analysis we can conclude that both the ad 
hoc procedure as well as the box potential fail in producing the exponent obtained for the more 
{\em generic} (regular) potentials.

\begin{figure}[t]
\begin{center}
   \includegraphics[width=.65\linewidth,clip]{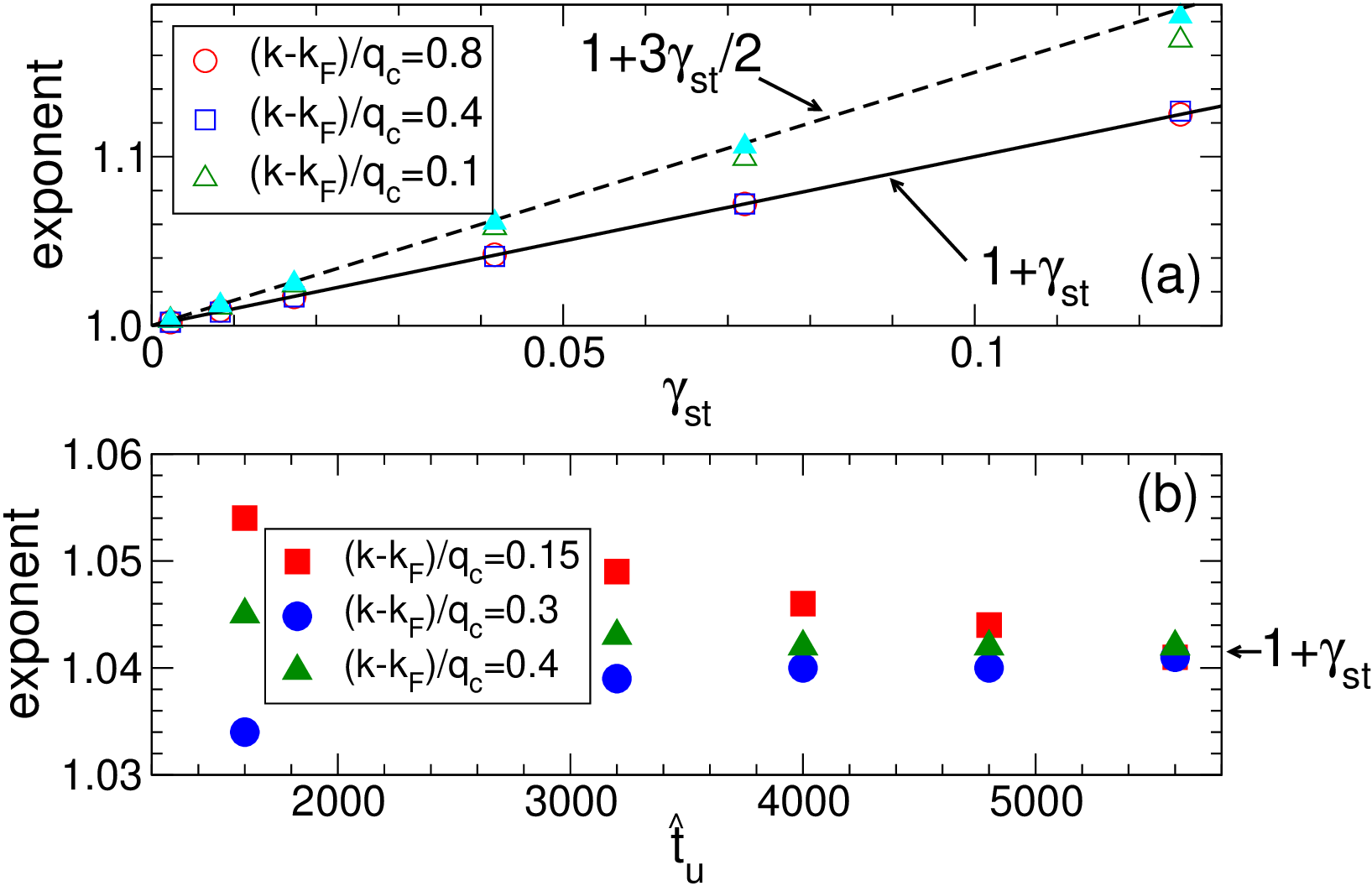}
   \caption{Exponents of the power-law decay of  $ | \Delta n (k,t)|$. (a) The open symbols 
   show the exponent as a function of $\gamma_{\rm st}$   
   for the Gaussian potential and the momenta as given in the legend. The filled light blue 
   triangles 
   are for the quartic potential with $\Delta \hat k=0.1$. The solid line shows 
   $1+\gamma_{\rm st}$ the dashed one the exponent $1+3 \gamma_{\rm st}/2$. (b) The filled symbols
   show the dependence of the exponent on the upper boundary of the time interval over 
   which the power law was fitted. The lower boundary is $\hat t=200$. The data were obtained 
   for the Gaussian potential with 
   $\hat v=0.5$, corresponding to $\gamma_{\rm st} \approx 0.0417$ and $\Delta \hat k = 0.15$, 
   $0.3$ as well as $0.4$.}
   \label{fig4}
\end{center}
\end{figure}

In Fig.~\ref{fig4}(a) we collected the data for the decay exponent of $ | \Delta n (k,t)|$ 
as a function of $\gamma_{\rm st}$ for the Gaussian potential from Figs.~\ref{fig3}(a) and (b) 
and added another set obtained for $\Delta \hat k=0.1$ (open symbols). 
For  small $\left| \Delta \hat k \right|$ the extracted exponents fall between the lines 
$1+3 \gamma_{\rm st}/2$ and $1+ \gamma_{\rm st}$ and one might be tempted to conclude 
that the exponent depends on $k-k_{\rm F}$. To further investigate this we studied the dependence 
of the extracted exponent on the time interval over which the power law was fitted. We 
increased the upper boundary $\hat t_{\rm u}$ at fixed lower boundary $\hat t_{\rm l}$.   
For the three regular potentials and all the $\Delta \hat k$ we studied we found that by 
increasing $\hat t_{\rm u}$ the exponent tends towards $1+ \gamma_{\rm st}$. In 
Fig.~\ref{fig4}(b) we show the $\hat t_{\rm u}$ dependence 
of the extracted exponent for the Gaussian potential with $\hat v=0.5$, corresponding 
to $\gamma_{\rm st} \approx 0.0417$, and the relativ  momenta $\Delta \hat k=0.15$, 
 $\Delta \hat k=0.3$, and $\Delta \hat k=0.4$. For increasing fitting range with 
lower boundary  $\hat t_{\rm l}=200$ 
the exponent has a clear tendency towards $1 + \gamma_{\rm st}$ and appears to saturate for 
times we can reach within our numerics. We conclude that our results for the regular 
potentials are consistent 
with the assumption of an {\em asymptotic} power-law exponent which 
is independently of $k-k_{\rm F}$  given by $1+ \gamma_{\rm st}$ (as long as $|k-k_{\rm F}| \neq 0$).

Figure~\ref{fig4}(a) additionally contains data for the exponent of the quartic potential at 
$\Delta \hat k=0.1$ (filled light blue triangles) extracted for the time interval 
$\hat t \in[100,2400]$. The exponents 
are even closer to the result of the ad hoc procedure $1 + 3 \gamma_{\rm st}/2$  
than the corresponding ones for the Gaussian potential at the same $\Delta \hat k$ (open green 
triangles). This can be understood as follows. First one realizes that it is the momentum dependence 
of the dispersion $\omega(k)$ and not the one of the prefactor $s^2(k) c^2(k)$ 
in Eq.~(\ref{GTL}) which dominates the long-time dynamics of the Green function. 
In the {\em ad hoc procedure} the nonlinear dispersion $\omega(q)$ is replaced by the linear 
one $\tilde v_{\rm F} |q|$ (see Eq.~(\ref{pfusch_2})). For small momenta the 
{\em quartic potential} deviates from its zero momentum value only to fourth order 
in $q/q_{\rm c}$. This implies that for small $|q|/q_c$, $\omega(q)$ deviates from 
the linear behavior to fourth order and over the momentum range $0 \leq |q|/q_{\rm c} \lessapprox 1$ 
the dispersion  relation of the ad hoc procedure and the quartic potential coincide. This 
has to be contrasted to the deviation from a linear dispersion for the exponential and 
Gaussian potentials which is of first and second order in $q/q_{\rm c}$, respectively. 
In this sense the quartic potential is the one considered closest to the ad hoc 
procedure.\footnote{The box potential implies a strictly 
linear dispersion up to $q_{\rm c}$. The asymptotic dynamics still shows a different exponent 
compared to the one of the ad hoc procedure as the discontinuity of the potential strongly affects the 
long-time dynamics; see above.} For small $\left| \Delta \hat k \right|$ 
this implies, that an {\em apparent} power-law decay with exponent 
$1 + 3 \gamma_{\rm st}/2$ is observable on intermediate times. Based on an analysis of the type 
discussed in the last paragraph we conclude that on {\em asymptotic} 
time scales the exponent will cross over to   $1 + \gamma_{\rm st}$. 
     
Similar to the cases of the ad hoc procedure and the box potential  also 
for the more regular potentials $ \Delta n (k,t) $
oscillates around zero (with a decaying amplitude). A 
Fourier analysis (not shown) clearly reveals that the (dimensionless) frequency of the oscillation at 
fixed $k-k_{\rm F}$ is given by $2 \hat \omega(k-k_{\rm F})$. This is consistent with the 
result for the ad hoc procedure for which $2 \hat \omega(k-k_{\rm F}) = 
2 \hat v_{\rm F}|\Delta \hat k |$ because of the linearization.    

A detailed account of the time dependence of $n(k,t)$ after a quench from one 
repulsive interaction to another one (obtained by Fourier transforming 
the Green function Eq.~(\ref{GTLWW1WW2})) can be obtained along the same lines. 
This is left for future work.

\section{Summary}
\label{summary}

We have studied the quench dynamics of the TL model focusing on the single-particle 
Green function and the fermionic momentum distribution function. Instead of using 
an ad hoc procedure to regularize momentum integrals in the ultraviolet we kept 
the momentum dependence of the two-particle interaction and considered different  
potentials. The steady-state momentum distribution function close to $k_{\rm F}$ 
is independently of the detailed form of the potential characterized by a 
power-law nonanalyticity with an exponent $\gamma_{\rm st}$ which 
depends via the LL parameter $K$ (or $K_{\rm i}$ and $K_{\rm f}$ when quenching 
between two repulsive interactions) on the strength 
of the interaction. We raised the question if this type of universality 
extends beyond the TL model to all models falling into the LL class in 
equilibrium. Importantly, $\gamma_{\rm st}$ differs from the exponent $\gamma_{\rm gs}$ 
characterizing the ground-state momentum distribution function at the same interaction. 
The ground-state and steady-state  momentum distribution functions differ at weak coupling by a 
characteristic factor of two, known from systems in which pre-thermalized states 
appear.   
  
Our analytical as well as numerical results for the asymptotic time dependence 
of $n(k,t)$ at small to intermediate interaction and $k-k_{\rm F} \neq 0$ are 
consistent with 
\begin{eqnarray}
 \Delta n(k,t)  \sim \frac{\sin{[ \hat \nu(k) \hat t]}}{\hat t^\xi} \; . 
\label{superres}
\end{eqnarray}
For the class of generic regular potentials  Eqs.~(\ref{gausspot})-(\ref{quartpot}) the frequency 
$\hat \nu(k)$ is given by $2 \hat \omega(k-k_{\rm F})$ and the power-law exponent by 
$\xi = 1 + \gamma_{\rm st}$. While the frequency depends on the details of the 
potential (the nonlinearity of the dispersion)
the exponent of this class of potentials turns out to be {\em universal} in the LL 
sense; it is solely determined by the potential at zero momentum transfer and thus the 
LL parameter $K$. Both, the nonanalytic box potential as well as the commonly employed 
ad hoc procedure fail in reproducing this universal behavior. For the box potential 
$\xi=1$ independent of the interaction strength and the (dimensionless) frequency is 
for all $\Delta \hat k$ given by $2 \hat v_{\rm F}$. In the ad hoc procedure 
we obtain  a larger decay exponent $\xi = 1 + 3 \gamma_{\rm st}/2$ and the frequency 
of the linearized dispersion $ \hat \nu(k) = 2 \hat v_{\rm F} |\Delta \hat k |$. At 
$k=k_{\rm F}$, $n(k,t)$ has a Fermi liquid-like jump of height $Z(t) \sim \hat 
t^{-\gamma_{\rm st}}$. This holds independent of the potential considered. 
The time scale $\hat t_{\rm p}$ on which the power-law decay Eq.~(\ref{superres}) is observable
at fixed $k-k_{\rm F}$ increases if $k$ approaches $k_{\rm F}$. If $\hat t \gg \hat t_{\rm p}$ 
is not fulfilled the asymptotic behavior Eq.~(\ref{superres}) might be superimposed or even 
hidden by additional contributions.           

In analogy to the posed question of LL-like universality in the steady state it would 
again be very interesting to investigate if some form of universality in the 
{\em time dependence} extends beyond the TL model studying the quench dynamics of other
 models falling into the LL universality class in equilibrium. Two obvious candidates would be 
the decay exponent $1 + \gamma_{\rm st}$ of $ | \Delta n(k,t) |$ found for regular potentials 
in the TL model and the exponent $\gamma_{\rm st}$ of the jump at $k_{\rm F}$. 

Studying the TL model with momentum dependent two-particle potentials we here performed the 
necessary first step towards a more detailed understanding of {\em universality in the quench 
dynamics of LLs.}

\vspace*{.5cm}

\noindent {\bf Acknowledgments:} We are grateful to Jean-S\'ebastien Caux, Fabian Essler, 
Christoph Karrasch, Stefan Kehrein, Dante Kennes, Florian Marquardt, Kurt Sch\"onhammer, and Nils Wentzell  
for fruitful discussions. This work was supported by the DFG via the Emmy-Noether program 
(D.S.) and FOR 912 (V.M.).

{}

\end{document}